\documentclass[conference,letterpaper]{IEEEtran}
\ifdefined\pdfminorversion
\fi
\usepackage[T1]{fontenc}
\usepackage{amsmath,amssymb}
\usepackage{graphicx}
\usepackage{afterpage}
\usepackage{cite}
\usepackage{url}
\usepackage{array,booktabs,tabularx}
\newcolumntype{Y}{>{\raggedright\arraybackslash}X}
\title{LAMP: LLM-Driven Analog Placement through Adaptive Tool Composition}
\author{
\IEEEauthorblockN{Zecheng Xu, Guanhua Chen, Yumao Wu, and Keren Zhu*}
\IEEEauthorblockA{
State Key Laboratory of ASIC and System, Fudan University, Shanghai, China
}
}

\begin{document}
\maketitle
\thispagestyle{empty}

\begin{abstract}
Analog placement requires circuit-specific trade-offs among area, matching, routability, and electrical performance. Balancing these trade-offs through predefined objectives and fixed placement flows remains difficult. In this paper, we propose an analog placement framework in which a large language model (LLM) acts as a layout engineer through composable placement components. These components support constraint management, placement and legalization, and lightweight evaluation of routability and electrical performance. Using multimodal checkpoints of the initial and evolving design, the LLM selects and combines components to construct and revise its own placement flow. It also adapts the placement hierarchy by reopening modules or partitioning and regrouping devices, allowing local layouts to respond to global design needs. An experience skill provides reusable design knowledge to guide these decisions. Evaluation on two SMIC 180 nm circuits shows that the framework reduces placement area and improves most measured post-layout performance metrics.
\end{abstract}


\newcommand{\queueperformancetable}{%
  \ifnum\value{page}<3\relax
    \afterpage{\queueperformancetable}%
  \else
\begin{table*}[!t] \centering \caption{Post-layout electrical performance and placement area on the SAR ADC and LDO.} \label{tab:experimental-results}\label{tab:performance} \small \begin{tabular}{@{}lrrrr@{}} \toprule \multicolumn{5}{c}{(a) SAR ADC} \\ \midrule Method & SNDR (dB) & ENOB (bit) & SFDR (dB) & \shortstack{Placement area\\($10^3\,\mu\mathrm{m}^2$)} \\ \midrule MAGICAL & 31.9218 & 5.0103 & 44.4624 & 116.532 \\ B*-tree & \textemdash{} & \textemdash{} & \textemdash{} & \textemdash{} \\ Ours w/o Experience Skill & 33.9596 & 5.3488 & 48.0722 & 108.624 \\ Ours & 34.3717 & 5.4172 & 48.1905 & 96.991 \\ \midrule \multicolumn{5}{c}{(b) LDO} \\ \midrule Method & $\Delta V_{\max}$ (mV) & PM ($^{\circ}$) & \shortstack{PSRR @ 1 MHz\\(dB)} & \shortstack{Placement area\\($10^3\,\mu\mathrm{m}^2$)} \\ \midrule MAGICAL & 1836.161 & 64.725 & $-3.4774$ & 261.824 \\ B*-tree & 1758.070 & 66.637 & $-5.2919$ & 195.957 \\ Ours w/o Experience Skill & 1631.327 & 68.554 & $-2.6453$ & 210.078 \\ Ours & 1664.220 & 63.950 & $-1.6081$ & 229.813 \\ \bottomrule \end{tabular} \end{table*}
  \fi
}
\afterpage{\queueperformancetable}


\section{Introduction}

Placement is central to analog layout design because device locations affect parasitics, matching, and routability \cite{ref1}. A good placement balances these requirements against area and performance, with priorities that vary across circuits. Layout Engineer rely on their own experience to make these trade-offs and revise layouts according to the result of parasitic extraction and post-layout simulation \cite{ref2}. This manual iteration makes layout design time-consuming.

Automated analog placement has evolved through three stages. Constraint-driven methods optimize area and wirelength under geometric requirements \cite{ref3,ref4}. Performance-driven methods use parasitic models, learned estimates, or post-layout simulation to guide the search \cite{ref5,ref6,ref26}. Integrated flows combine netlist analysis, device generation, placement, and routing \cite{ref7,ref8}, with hierarchical optimization coordinating module layouts \cite{ref25,ref9}. However, predefined objectives may not fully capture circuit function or intermodule interactions, and fixed settings are difficult to reuse across circuits \cite{ref10}. Repeating full routing, extraction, and simulation is costly \cite{ref2}, and converting performance feedback into specific placement changes remains difficult.


\begin{figure}[!t]
\centering
\includegraphics[width=\columnwidth]{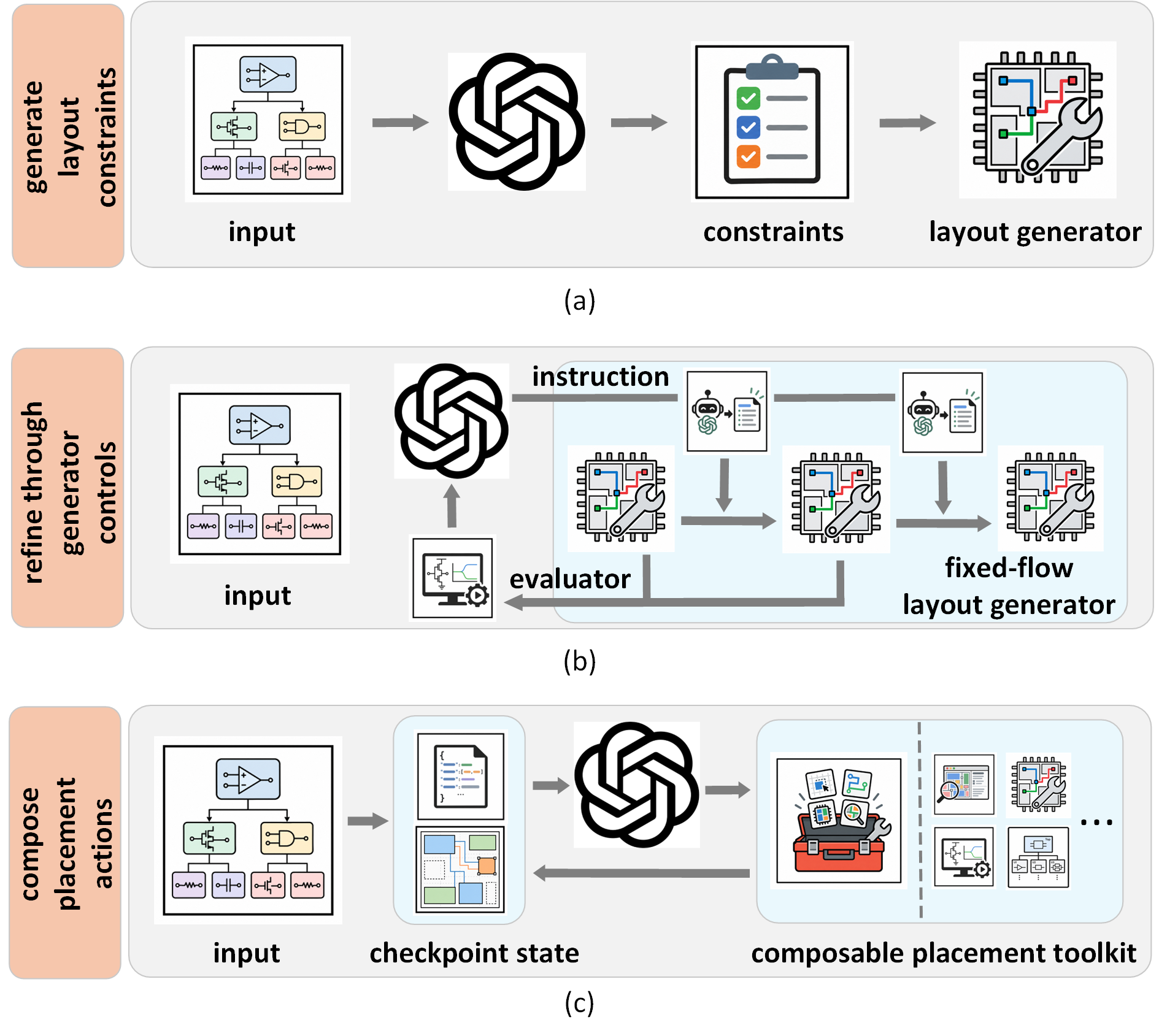}
\caption{Three ways to use an LLM in layout design: (a) generating layout constraints, (b) refining a fixed-flow layout generator through feedback, and (c) composing placement actions using the current checkpoint and a placement toolkit.}
\label{fig:llm-comparison}
\end{figure}

\begin{figure*}[!t]
\centering
\includegraphics[width=\textwidth]{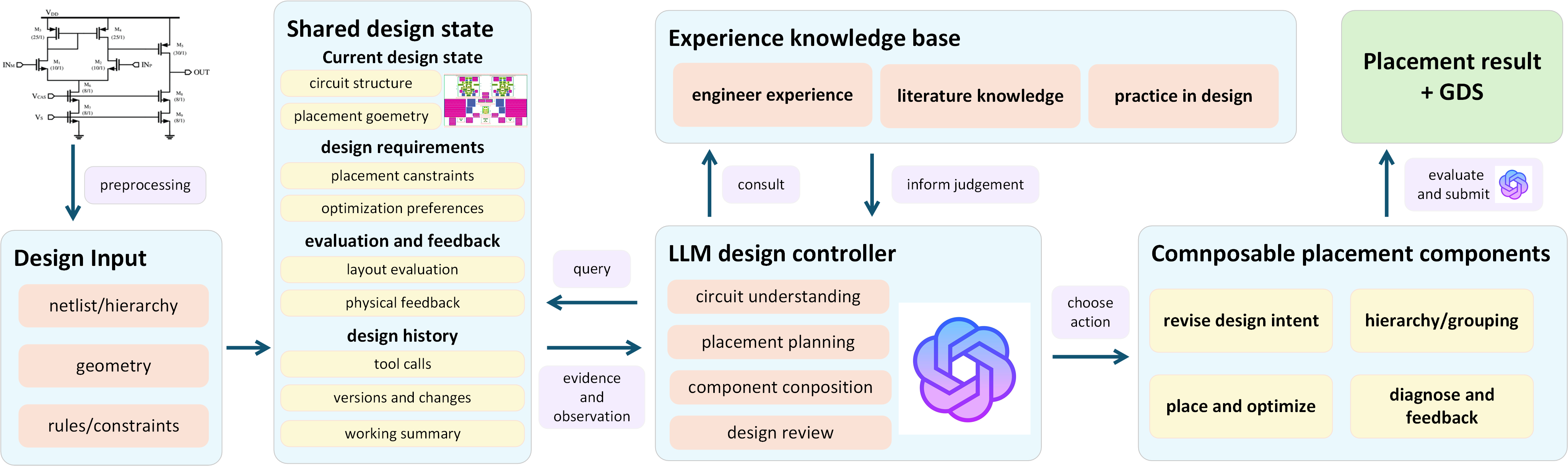}
\caption{Overview of the proposed framework. The LLM observes the shared design state and selects composable placement components. Execution results and physical feedback inform subsequent decisions, while the experience skill provides design knowledge. The selected placement is exported after checking hard placement constraints.}
\label{fig:framework}
\end{figure*}

Recent work uses large language models (LLMs) to translate natural-language requests into layout commands \cite{ref11,ref13} or directly predict device placements \cite{ref23}. Other approaches generate layout constraints \cite{ref21,ref24}, connect circuit interpretation and sizing to layout generation \cite{ref14,ref22}, or refine layout parameters using post-layout simulation feedback \cite{ref12,ref2}. However, in constraint- and parameter-based methods, the underlying engine still determines how the LLM's guidance becomes placement geometry \cite{ref12,ref21,ref2}. These interfaces can limit control over placement operations as the layout evolves. Meanwhile, advances in long-context handling, image understanding, and tool use support a broader role for LLMs in placement. Fig.~\ref{fig:llm-comparison} summarizes how existing approaches incorporate LLMs into analog placement.

In this paper, we introduce composable placement components (CPCs) for circuit observation, constraint specification, and placement optimization. The LLM selects and invokes these components throughout placement, adapting the flow based on its global understanding of the circuit and layout. Checkpoints summarize the current state, while rollback restores earlier placements. These mechanisms support more frequent and effective LLM intervention, improving placement efficiency and quality. Our main contributions are as follows:

\begin{itemize}
\item We replace a predefined layout generator with independently callable, composable placement components. The LLM combines these components to construct a placement flow suited to the design.
\item We introduce an iterative design flow built around compact multimodal checkpoints, which preserve design context across decisions. Checkpoint information enables LLM inspection and comparison, serving as a foundation for subsequent decisions and component routing.
\item We make the placement hierarchy adaptable during optimization. The LLM can reopen existing modules or partition and regroup devices according to function and connectivity. This flexibility allows local layouts to be revised in response to global design needs.
\end{itemize}

\section{Framework}

The framework proposed in this paper connects circuit interpretation, placement construction, and physical feedback through composable placement components. As shown in Fig.~\ref{fig:framework}, the design inputs initialize the state, which the LLM uses to understand the circuit and plan its placement actions. The LLM then consults the experience skill and invokes CPCs to construct and evaluate candidate layouts. Execution results and physical feedback update the shared state, providing the input for the next round of decisions. Compared with approaches that primarily use an LLM as an inspector or decider, the model in our framework plays a role closer to that of a layout engineer.


\subsection{Composable Placement Components}

We introduce composable placement components to address the difficulty of adapting conventional placement methods to circuits with different structures and constraints. The functions of CPCs include circuit inspection, layout construction, and physical evaluation. Each component is independently callable and exposes the information or controls needed for a specific placement task. By controlling these callable components rather than merely adjusting solver parameters, the LLM can precisely translate layout intent into the physical layout.

Besides directly assigning device coordinates, the LLM can also express an intended arrangement through constraints. Symmetry, spacing, and grouping requirements describe how devices should be organized, while net weights and signal or current path guidance emphasize important connections. For example, a spacing constraint may be inserted between two device groups to improve routability or pin access during subsequent placement. Four solving components search for placements under these requirements: B*-tree \cite{ref20}, nonlinear programming (NLP), CP-SAT, and one-dimensional legalization. Hard input constraints remain fixed and define acceptable solutions, while soft preferences allow trade-offs among competing objectives. Meanwhile, these solver components are invoked for further refinement and legalization to obtain optimized legal placements. 

The components also include several physical feedback tools. RUDY congestion estimation \cite{ref29} and pin access analysis assess routing difficulty, while post-layout simulation evaluates circuit performance with extracted interconnect parasitics. Reports from the lightweight analyses associate routing issues with affected devices or nets, providing targets for revision. These checks can be executed dynamically between placement changes whenever the LLM deems it necessary, allowing potential issues to be identified and resolved early before they compound. In summary, The LLM can combine direct edits, numerical search, and feedback in different orders, exploring alternatives beyond a fixed solver sequence.

\subsection{Global Circuit Understanding and Placement Control}

Conventional placement algorithms usually optimize predefined objectives such as area and half-perimeter wirelength (HPWL) under specified constraints. The LLM can also identify circuit requirements that are difficult to quantify, using its understanding of functional roles and connections across the full design. It interprets the current layout and physical feedback in this context, weighs competing needs, and determines what the next placement step should achieve. Then it selects the placement scope and invokes the appropriate components.

After component calls execute, the framework updates the shared design state with the resulting geometry, guidance, and evaluations. It rebuilds a compact checkpoint of the current placement and requirements, combining it with execution results and layout images to form the input for the next decision. Within this input, structured summaries report active constraints and measured quality, while images show spatial relationships and available physical heatmaps. The LLM assesses the updated placement against its intended goals and examines unresolved issues before deciding how to proceed. When deeper detail is needed, the full circuit data and original requirements remain available through inspection components.

Saved versions preserve both placement and guidance, allowing rollback after an unsuccessful trial. The LLM can compare their evaluations and inspect changes in geometry and requirements before deciding which candidate to retain. A working summary retains the reasoning behind earlier decisions and unresolved issues across rounds. Once the LLM considers a placement satisfactory, the framework checks all hard placement constraints. The final placement and GDS are determined by the LLM and can be submitted only after those checks pass.

\subsection{Adaptive Hierarchical Placement}

Hierarchy reduces placement complexity by optimizing subcircuits separately and assembling their layouts \cite{ref25,ref9}. A completed module is treated as a rigid object during assembly, with its shape and relative pin locations fixed. However, fixing local layouts can exclude useful arrangements at the parent level. A compact child may have unfavorable pin locations for external connections, turning a good local solution into a poor choice for the assembled circuit.

\begin{figure}[!ht]
\centering
\includegraphics[width=\columnwidth]{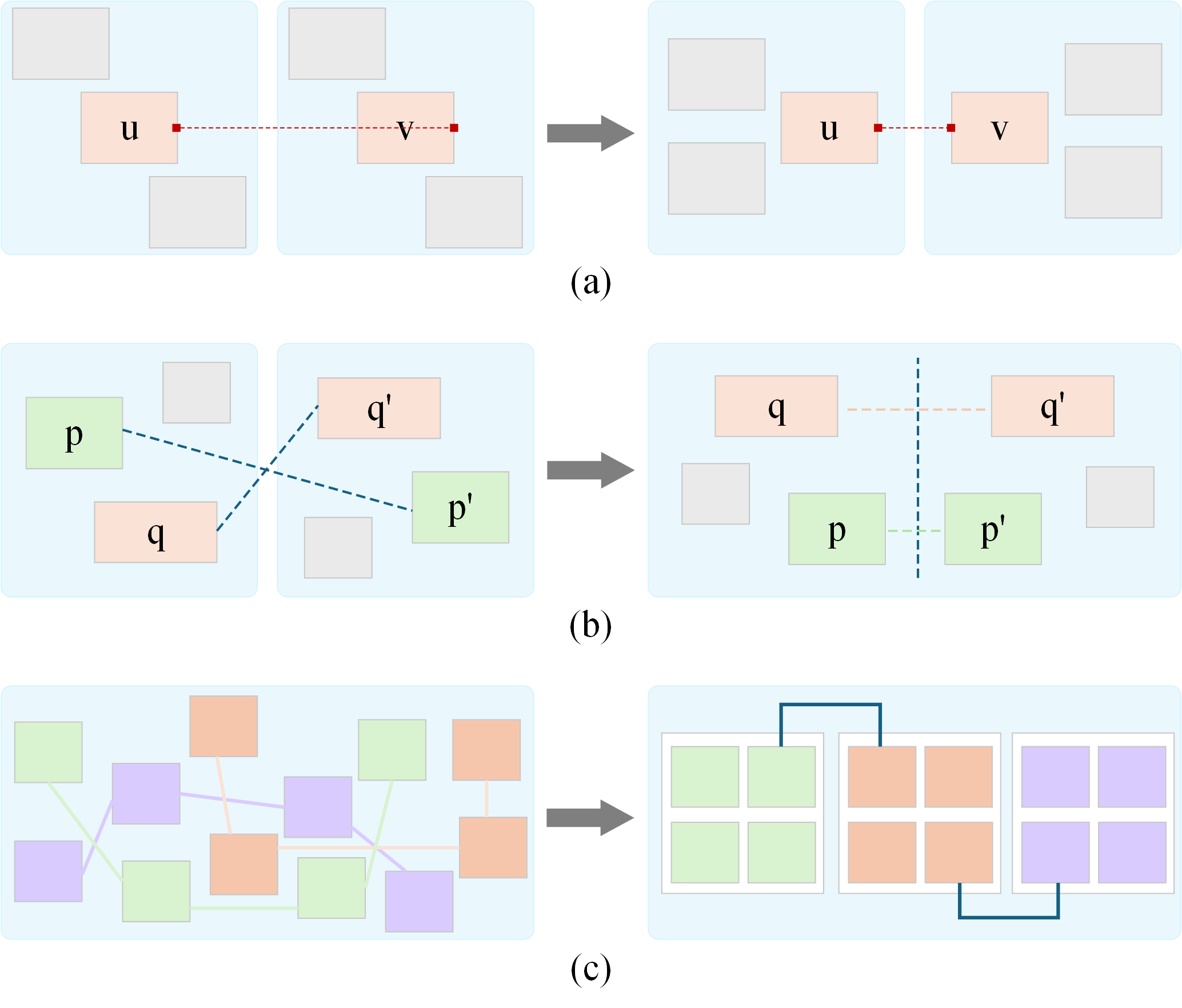}
\caption{Three ways to adapt placement organization: (a) reopening a module to improve its interfaces, (b) flattening and regrouping devices for matching across module boundaries, and (c) partitioning a flat scope using functional and connectivity requirements.}
\label{fig:hierarchy}
\end{figure}

The LLM can revisit these decisions using circuit semantics and physical feedback. In Fig.~\ref{fig:hierarchy}(a), it reopens a child whose interface limits the parent, adjusts its internal placement, and reassesses the updated module after assembly. The local solve accounts for connections outside the module through available external endpoints and interface preferences. When matching requirements or connectivity favor joint optimization across module boundaries, the LLM can flatten the relevant scope and regroup the devices, as in Fig.~\ref{fig:hierarchy}(b).

For a large flat circuit, the LLM instead uses circuit function and connectivity to define smaller physical groups. It identifies devices that should remain together and important nets that guide hypergraph partitioning \cite{ref27,ref28}. The partitioner balances group area while discouraging cuts through important connections. The LLM can revise these proposed groups before committing to local placement. The groups are placed separately and assembled, as shown in Fig.~\ref{fig:hierarchy}(c). The LLM evaluates the intergroup connections and can refine the assembled layout to seek a better overall arrangement. 

\subsection{Experience Skill}

The experience skill serves as a knowledge base with three sources: engineering experience, knowledge from the literature \cite{ref17,ref18,ref19}, and lessons derived from layout iterations. The literature connects device arrangement to matching accuracy and interconnect parasitics. Each entry records the physical rationale for a design choice and the conditions under which it applies. The LLM checks these conditions against the current circuit and feedback before using an entry to guide its next action.

We also propose an update process that compares the intended effect of a placement change with physical evaluation or simulation results. The LLM analyzes successes and failures, then extracts general lessons while removing design-specific details. As new results support or challenge earlier lessons, the LLM revises their applicable conditions or consolidates consistent findings, so that later decisions can build on both accumulated experience and corrections to earlier judgments.

\section{Experimental Results}

\subsection{Experimental Setup}

We evaluate the proposed framework on an 8-bit SAR ADC and a low-dropout regulator (LDO) in SMIC 180~nm CMOS. The hierarchical SAR ADC contains 1544 devices and operates at 1.8~V. The flat LDO contains 202 devices and provides a 3.3~V output. Inputs are exported from schematics and PDK PCells using Cadence Virtuoso. The placements are routed using the same adapted MAGICAL anaroute backend and the technology rules for that circuit. We use Calibre for DRC and LVS checks and RC parasitic extraction. Post-layout simulation is performed using Spectre with the extracted DUT in the corresponding circuit testbench. All electrical measurements use the TT model corner at $27\,^{\circ}\mathrm{C}$.

We compare the proposed framework with three baselines. \emph{MAGICAL} uses the hierarchical analytical placement flow of the open-source layout generator \cite{ref8}. \emph{B*-tree} is our implementation of hierarchical simulated annealing using the B*-tree representation \cite{ref20,ref4}. \emph{Ours w/o Experience Skill} removes the experience skill while retaining component instructions. Both LLM variants use GPT-6.1-sol with identical placement components and feedback interfaces. For a fair comparison, experience evolution uses only ADC1 and ADC2 from MAGICAL, and the skill is frozen before evaluation on the SAR ADC and LDO.

For the SAR ADC, we use deterministic transient simulations with a 40~MHz clock and a 117.1875~kHz differential sinusoidal input. A 1024-point coherent FFT of the reconstructed output provides SNDR, ENOB, and SFDR. SNDR measures signal quality relative to noise and distortion, ENOB expresses this quality as an effective bit resolution, and SFDR measures how far the strongest spur lies below the fundamental. For the LDO, we use a 10~pF output capacitor and a 1--100--1~mA load pulse with $1\,\mu\mathrm{s}$ rise and fall times. The maximum absolute output deviation from 3.3~V over $[30,100]\,\mu\mathrm{s}$ characterizes the transient response. To assess main-loop stability, we measure phase margin (PM) using Spectre STB with a probe between VOUT and feedback resistor R7 and report the smaller value at 1 and 100~mA. PSRR at 1~MHz, measured at a static load of 100~mA, quantifies supply ripple rejection. Area denotes the bounding-box area of all placed cells before routing.

\subsection{Results and Ablation Analysis}

Table~\ref{tab:experimental-results} presents the experimental results. The proposed framework successfully completes placement and routing for both designs, supporting its robustness and scalability across the evaluated cases.

For the hierarchical SAR ADC, \emph{Ours} reduces area by 16.77\% relative to MAGICAL while improving SNDR by 2.45~dB and SFDR by 3.73~dB. These simultaneous gains indicate that the framework can find better arrangements within the circuit hierarchy while maintaining favorable electrical performance after routing. \emph{Ours w/o Experience Skill} also improves electrical performance and area over MAGICAL. Enabling the experience skill further reduces area by 10.71\% and shows small improvements in all three electrical metrics, validating the necessity of the collected knowledge and accumulated experience.

For the flat 202-device LDO, \emph{Ours} reduces maximum output deviation by 9.36\% and area by 12.23\% relative to MAGICAL, while improving PSRR by 1.87~dB. It achieves the highest PSRR among the four methods, whereas B*-tree produces the smallest area but the weakest supply rejection. The framework thus offers a competitive balance between compactness and electrical performance on this flat design. Enabling the experience skill improves PSRR by another 1.04~dB, but increases output deviation and area and reduces PM. For designs with many cells, our method can therefore provide a practical solution on the Pareto frontier of the evaluated placements, balancing area and electrical performance.

\section{Conclusion}

This paper presented an LLM-driven analog placement framework built on composable placement components. Guided by circuit understanding and physical feedback, the LLM combines circuit inspection, constraint specification, direct edits, and numerical methods into an adaptive placement flow. Multimodal checkpoints preserve design context and enable llm functions like comparison and rollback. Adaptive hierarchy supports reopening placed modules, regrouping devices across module boundaries, and partitioning flat circuits according to function and connectivity. Compared with MAGICAL, the framework reduces SAR ADC placement area by 16.77\% and improves post-layout SNDR by 2.45~dB, while reducing LDO placement area and maximum output deviation by 12.23\% and 9.36\%, respectively. These results demonstrate favorable trade-offs between area and electrical performance across both designs, with particularly strong gains on the hierarchical designs. 
\clearpage

\end{document}